\documentclass[prd,aps,floats,superscriptaddress,preprint]{revtex4}
\usepackage[paperwidth=210mm,paperheight=297mm,centering,hmargin=1.5cm,vmargin=1.5cm]{geometry}

\usepackage{fouriernc}

\usepackage[T1]{fontenc}
\renewcommand*\ttdefault{cmtt}
\usepackage{cabin}
\usepackage{savesym}
\savesymbol{mathbb}
\savesymbol{mathcal}
\savesymbol{mathfrak}
\savesymbol{mathscr}
\savesymbol{circledS}

\everymath{\displaystyle}

\usepackage{graphicx}
\usepackage{float}
\usepackage{physics}

\usepackage{gensymb}
\usepackage{slashed}

\usepackage{tabularx}

\newcommand{\mathsym}[1]{{}}

\clearpage 
\cleardoublepage

\newcommand{\m}{\mu}

\newcommand{\aG}{\alpha}
\newcommand{\bG}{\beta}

\newcommand{\LG}{\Lambda}
\newcommand{\lG}{\lambda}

\usepackage{gensymb}
\usepackage{slashed}

\usepackage{tabularx}
\usepackage{graphicx, amsmath, amssymb, graphicx, float, hyperref}
\usepackage[toc,page]{appendix}
\usepackage[usenames,dvipsnames,svgnames]{xcolor}

\usepackage{braket}

\usepackage{physics}

\usepackage{amsmath}

\usepackage{amssymb}

\usepackage{upquote}

\usepackage{mathtools}

\clearpage 
\cleardoublepage

\newcommand{\be}{\begin{equation}}
\newcommand{\ee}{\end{equation}}
\newcommand{\bea}{\begin{eqnarray}}
\newcommand{\eea}{\end{eqnarray}}

\begin{document}
	
	\setlength{\unitlength}{1mm}
	
	\title{\textsc{Thin-shell instanton tunneling: something-to-something or nothing-to-something?}\footnote{Proceedings for \textit{The 2nd LeCosPA Symposium: Everything about Gravity, Celebrating the Centenary
				of Einstein's General Relativity}. Talk given by Yao-Chieh Hu, on December 17, 2015, Taipei, Taiwan.}}

	\author{Pisin Chen}
	\email{pisinchen@phys.ntu.edu.tw}
	\affiliation{Department~of~Physics, National~Taiwan~University, Taipei~10617, Taiwan, R.O.C.\\}
	\affiliation{Leung~Center~for~Cosmology~and~Particle~Astrophysics, National~Taiwan~University, Taipei~10617, Taiwan, R.O.C.\\}
	\affiliation{Graduate~Institute~of~Astrophysics, National~Taiwan~University, Taipei~10617, Taiwan, R.O.C.\\}
	\affiliation{Kavli~Institute~for~Particle~Astrophysics~and~Cosmology, SLAC~National~Accelerator~Laboratory, Stanford~University, Stanford, CA~94305, U.S.A.\\}
	
	\author{Yao-Chieh Hu}
	\email{r04244003@ntu.edu.tw}
	\affiliation{Department~of~Physics, National~Taiwan~University, Taipei~10617, Taiwan, R.O.C.\\}
	\affiliation{Leung~Center~for~Cosmology~and~Particle~Astrophysics, National~Taiwan~University, Taipei~10617, Taiwan, R.O.C.\\}
	\affiliation{Graduate~Institute~of~Astrophysics, National~Taiwan~University, Taipei~10617, Taiwan, R.O.C.\\}
	
	\author{Dong-han Yeom}
	\email{innocent.yeom@gmail.com}
	\affiliation{Leung~Center~for~Cosmology~and~Particle~Astrophysics, National~Taiwan~University, Taipei~10617, Taiwan, R.O.C.\\}

	\begin{abstract}
		There exist two interpretations of instantons in the literature. The first interpretation regards instanton as divider between the initial and final hypersurfaces. The Coleman-De Luccia instanton is one such an example. The second interpretation, proposed by Brown and Weinberg, considers instanton as connector between the initial and final hypersurfaces. In this proceedings, we try to suitably and intuitively argue that these two interpretations are complementary to each other under certain conditions. Furthermore, we demonstrate that the decay rate obtained from the Euclidean treatment and the Hamiltonian treatment both are consistent with each other, which may help to dissolve some concerns about the validity of regularization technique employed in the treatment of the cusp singularity of instantons. Based on these, we argue that instantons can be a sensible tool to address the information loss problem. 
	\end{abstract}
	
	\maketitle

	\section{Introduction}
	
	Instanton is an interesting and extremely important object in quantum field theory, which describes nonperturbative phenomenon of a system. In this article we summarize our recent work \cite{PYD} on interpretations of instantons in the Euclidean path-integral approach, and in particular, its connection with black hole physics.
	
	By interpretation, we actually refer to the technical issue of how an instanton, which begins with the Euclidean spacetime, can be analytically continued to the eventual Lorentzian one. In the literature, it is known that there are two interpretations of instanton that can be summarized as follows. For simplicity, we shall refer to these two interpretations as \textit{Nothing-to-Something} and \textit{Something-to-Something}, where
	
	\begin{itemize}
		\item[--] \textit{Nothing-to-Something interpretation}: the initial and final hypersurfaces are separated by an instanton \cite{Coleman&De Luccia}.
		\item[--] \textit{Something-to-Something interpretation}: the initial and final hypersurfaces are connected by an instanton \cite{Brown&Weinberg}.
	\end{itemize}
	
	So far the investigation of instanton solutions in the models of Coleman and De Luccia\cite{Coleman&De Luccia} and Brown and Weinberg\cite{Brown&Weinberg} have been based on O(4) symmetry. In our work, we go further to study with the spherical symmetry that can include objects like black holes. 
	
	Besides the issues of interpretations, both Euclidean and Hamiltonian approaches have been invoked in the analysis of the something-to-something interpretation. On the contrary, for the nothing-to-something interpretation, there was no article that used the Hamiltonian treatment to calculate the decay rate. To make sure the consistency between two interpretations, it is worthwhile to check that the Hamiltonian approach also works well for the nothing-to-something interpretation. We therefore test the consistency of the Euclidean treatment by computing the decay rate from the Hamiltonian treatment.

	\section{Two interpretations---de Sitter space}
	
	Here we list the pros and cons of the two interpretations with \(O(4)\) symmetry. 
	\begin{figure}
		\begin{center}
			\includegraphics[width=5in]{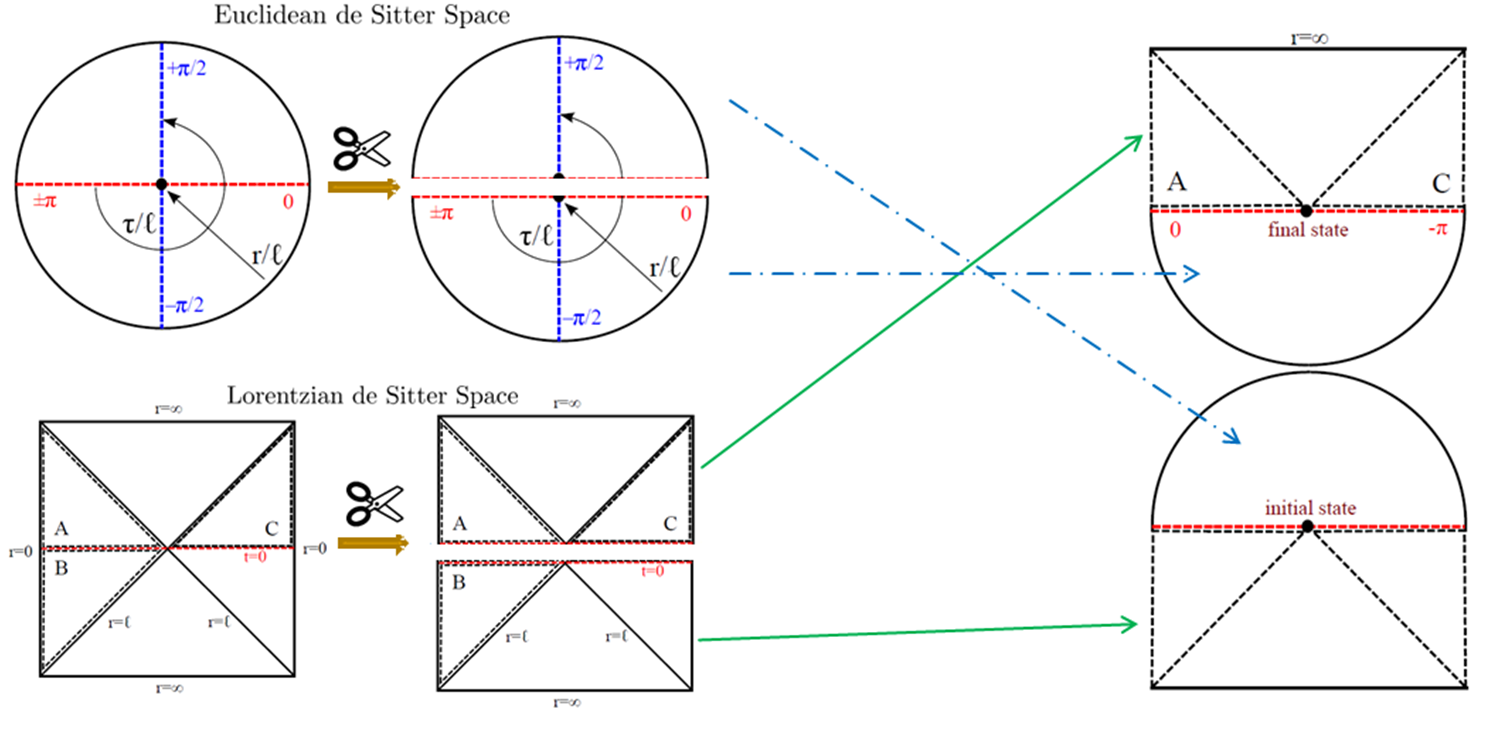}
			\caption{Pictorial illustration for the nothing-to-something interpretation. One can easily see that the glued manifold is smoothly connected and maximally extended but contains some unphysical region beyond the Hubble radius.}
			\label{aba:N-S}
		\end{center}
		\begin{center}
			\includegraphics[width=5in]{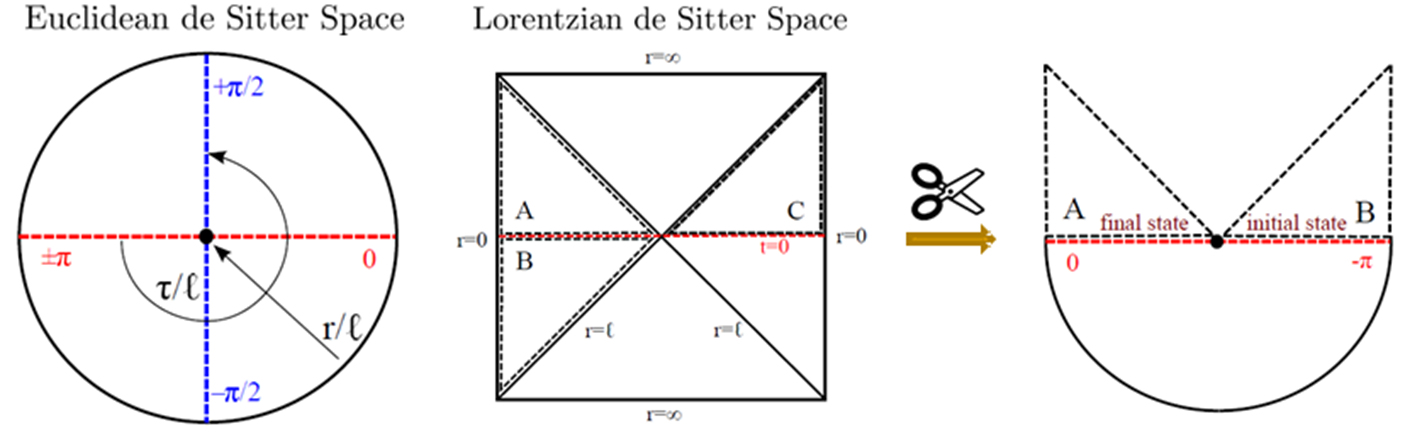}
			\caption{Pictorial illustration for the something-to-something interpretation. It was shown by Brown and Weinberg \cite{Brown&Weinberg} that once we identify the initial hypersurface with the final hypersurface between Euclidean and Lorentzian manifolds, a consistent interpretation is possible.}
			\label{aba:S-S}
		\end{center}
	\end{figure}
	
	\subsection{Nothing-to-Something}
	
	As we mentioned before, the nothing-to-something interpretation is the case where the initial and final hypersurfaces are separated by the instanton solution (Fig.~\ref{aba:N-S}). This is the traditional interpretation that smoothly glues and maximally extends manifold, which is of course mathematically natural. In this interpretation, however, the manifold contains a region that lies beyond the cosmological horizon and hence inaccessible. 
	
	\subsection{Something-to-something}
	For the something-to-something interpretation, in which the initial and final hypersurfaces are connected by the instanton solution, the tunneling process takes place in one causal patch. This interpretation is therefore more physically intuitive. However, there are still some mathematically technical issues; the manifold is not maximally extended and it is difficult to apply it to some important cases such as including the anti-de Sitter space.
	
	\section{Two interpretations---spherical symmetry}
	
	For our own interests in black hole physics, we move to the spherical symmetry and testify whether it is possible to interpret in both ways. In this section, we will briefly review the dynamics of a thin-shell bubble. Then, we will introduce both Euclidean and Hamiltonian methods in order to describe the tunneling process.
	
	\subsection{Dynamics of thin-shell bubbles}
	Let us consider a true vacuum bubble 
	of the scalar field \(\phi\) with the following action\cite{PYD}: 
	\bea
	S=\int_M\sqrt{-g} d^4x\left[\frac{\mathcal{R}}{16\pi}-\frac{1}{2}\nabla^\mu\phi \nabla_\mu\phi-U(\phi)\right]
	+\int_{\partial M}\sqrt{-h}d^3x\left[\frac{\mathcal{K}-\mathcal{K}_0}{8\pi}\right],
	\eea
	where $\mathcal{R}$ is the Ricci scalar, $\phi$ is a scalar field, $U(\phi)$ is the potential of the scalar field, and $\mathcal{K}$ and $\mathcal{K}_{0}$ are Gibbons-Hawking boundary terms for a given metric and the Minkowski metric, respectively.
	With the spherical symmetry, one can easily obtain the equation of motion of the bubble wall by using the thin-shell approximation: after straightforward calculations, we obtain the following form of the equation\cite{PYD}: 
	\bea\label{EOM}
	\dot{r}^2+V(r)=0,
	\eea
	where the effective potential \(V(r)\) is a function of several physical quantities, such as the mass and the surface energy density of the black hole. 
	
	\subsection{Two approaches toward tunneling process}
	
\begin{figure}
		\begin{center}
			\parbox{3in}{\includegraphics[width=3in]{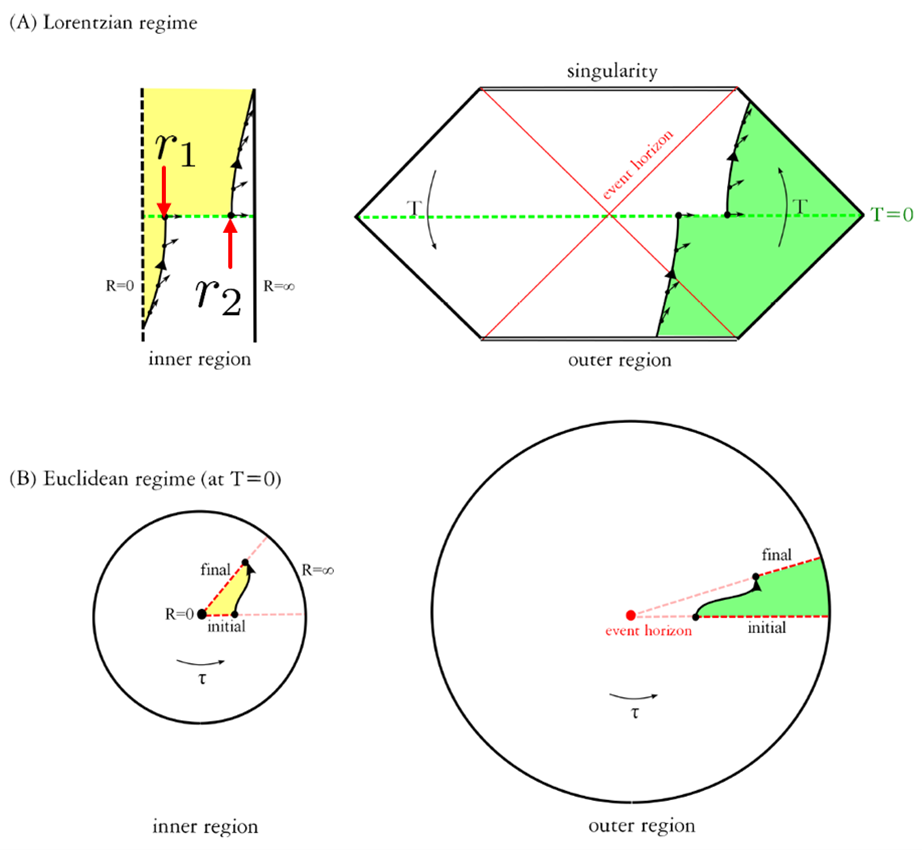}}
			\hspace*{4pt}
			\parbox{3in}{\includegraphics[width=3in]{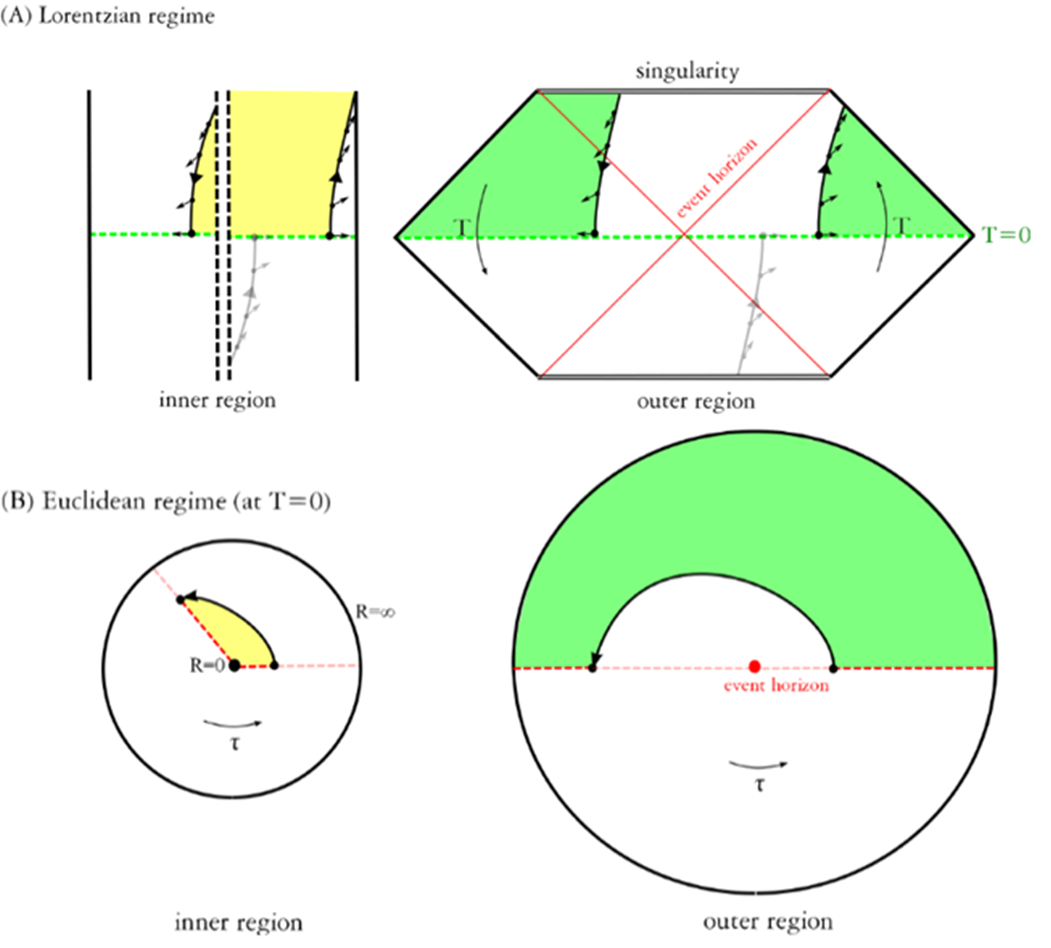}}
			\caption{In the case of the true vacuum decay problem in the Minkowski background, the inside region has the anti-de Sitter geometry and the outside region has the Schwarzschild geometry, where where the something-to-something interpretation is on the left and the nothing-to-something interpretation is on the right.}
			\label{ns}
		\end{center}
	\end{figure}

	There are two approaches to treat the tunneling process. The first one is the Euclidean method, for example Farhi-Guth-Guven method \cite{FGG} that relies on the Euclidean path integral. The second one is the Hamiltonian method proposed by Fischler-Morgan-Polchinski \cite{FMP}, where the integration is performed over the constant-time hypersurfaces. In the literature, both methods are adopt under the something-to-something interpretation. However, both methods can be interpreted as nothing-to-something as long as the Euclidean time period is properly chosen, that is \(\tau_{r_1\rightarrow r_2}=8\pi M\) (Fig.~\ref{ns}). We can use the \textit{true vacuum decay} as a demonstration as follows.
	
	For the nothing-to-something interpretation, though the Euclidean method is verified by Gregory-Moss-Withers \cite{GMW}, the Hamiltonian method has not been clarified. Therefore, we perform a consistency check as follows.
	
	\section{Consistency check}
	
	Both Farhi-Guth-Guven (FGG) \cite{FGG} and Fischler-Morgan-Polchinski (FMP) \cite{FMP} tunnelings give the same decay rates. These methods are therefore equally valid. A question then naturally arises--- Can one also find a coherent decay rate between Euclidean and Hamiltonian approaches for nothing-to-something interpretation?
	\begin{table}
		\tablename{: Comparison table of our consistency checks.}
		{\begin{tabular}{@{}cccc@{}}
				\toprule
				& Something-to-Something & Nothing-to-Something 
				\\\colrule
				Euclidean Approach (FGG) & FGG\cite{FGG} & GMW\cite{GMW} \\
				Hamiltonian Approach (FMP) & FMP\cite{FMP} & CHY\cite{PYD} \\
				
			\end{tabular}}
		\end{table}

		\subsection{Decay rates in the Euclidean approach}
		In the Euclidean approach, the decay rate is given as
		\bea
		\Gamma\propto e^{-2B},
		\eea
		where \(B=S_{\mathrm{E}}(\mathrm{instanton})-S_{\mathrm{E}}(\mathrm{background})\) is the tunneling exponent and the Euclidean action takes the form
		\bea
		S_{\mathrm{E}}=-\int_M\sqrt{+g} d^4x\left[\frac{\mathcal{R}}{16\pi}-\frac{1}{2}\nabla^\mu\phi \nabla_\mu\phi-U(\phi)\right]
		-\int_{\partial M}\sqrt{+h}d^3x\left[\frac{\mathcal{K}-\mathcal{K}_0}{8\pi}\right].
		\eea
		The decay rate with the nothing-to-something interpretation has been derived by Gregory-Moss-Withers \cite{GMW}, with
		\bea\label{tunnelingexponent}
		2B=\frac{\mathcal{A}_i-\mathcal{A}_f}{4}+2\int_{r_1}^{r_2}dr r 
		\bigg|\cos^{-1}\bigg(\frac{f_++f_--16\pi^2\sigma^2r^2}{2\sqrt{f_+f_-}}\bigg) \bigg|,
		\eea
		where \(f_+ \equiv 1-2M/r\), \(f_- \equiv 1+ r^2/l^2\), and $\mathcal{A}_{i,f}$ are areas of horizons. Note that the first part of the right hand side corresponds to the entropy term which came from the regularization technique around the cusp singularity.
		
		\subsection{Decay rates in the Hamiltonian approach}
\begin{figure}
			\begin{center}
				\includegraphics[width=3in]{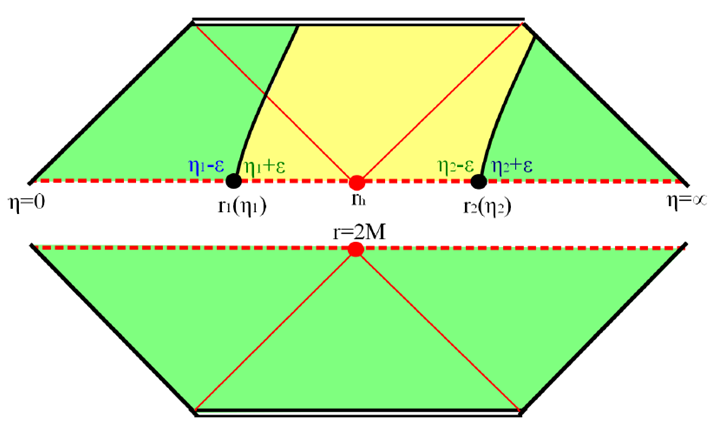}
				\caption{Integration taken over spacelike hypersurfaces.}
				\label{aba:int}
			\end{center}
		\end{figure}

		In the Hamiltonian approach, the decay rate is given by \cite{FMP}
		\bea
		\Gamma\propto e^{2i(\Sigma_f-\Sigma_i)},
		\eea
		where \(\Sigma_f-\Sigma_i\) is the difference between the integrals on initial and final spacelike hypersurfaces (Fig.~\ref{aba:int}). Following the nothing-to-something interpretation, we \cite{PYD} computed the action integration and we have shown that the integration takes the form
		\bea
		\left[\int_{0}^{\eta_1-\epsilon}d\eta(\dots)+\int_{\eta_1+\epsilon}^{\eta_2-\epsilon}d\eta(\dots)
		+\int_{\eta_2-\epsilon}^{\eta_2+\epsilon}d\eta(\dots) \right]
		+\left[\int_{\eta_1-\epsilon}^{\eta_1+\epsilon}+\int_{\eta_2-\epsilon}^{\eta_2+\epsilon}\right],
		\eea 
		where the first parenthesis was referred to as the \textit{volume-integration} and the second the \textit{shell-integration}.
		
		We have showed that\cite{PYD}
		\bea
		\left[\int_{0}^{\eta_1-\epsilon}d\eta(\dots)+\int_{\eta_1+\epsilon}^{\eta_2-\epsilon}d\eta(\dots)
		+\int_{\eta_2-\epsilon}^{\eta_2+\epsilon}d\eta(\dots)+\right] = \frac{\mathcal{A}_i-\mathcal{A}_f}{4}
		\eea 
		and 
		\bea
		\left[\int_{\eta_1-\epsilon}^{\eta_1+\epsilon}+\int_{\eta_2-\epsilon}^{\eta_2+\epsilon}\right] = 
		2\int_{r_1}^{r_2}dr r 
		\bigg|\cos^{-1}\bigg(\frac{f_++f_--16\pi^2\sigma^2r^2}{2\sqrt{f_+f_-}}\bigg) \bigg|.
		\eea
		Detail calculations can be found in \hyperref[app]{\textbf{APPENDIX:A}}.
		Thus, we have demonstrated that not only the Euclidean but also the Hamiltonian method work well for the nothing-to-something interpretation. Therefore, we can safely say that these two interpretations are complementary with each other.
		
		\section{Conclusion}
		
		We have shown that the two interpretations on instantons are complementary each other under some constraint conditions. In addition, by checking consistency between the Euclidean method and the Hamiltonian method, we have verified that the regularization technique of Gregory-Moss-Withers is valid.
		
		\section*{Acknowledgment}
		YH would like to thank Hsu-Wen Chiang and Chun-Yen Chen for beneficial discussions. PC and YH are supported by Taiwan's Ministry of Science and Technology (MOST 104-2112-M-002-024-MY3). DY is supported by Leung Center for Cosmology and Particle Astrophysics (LeCosPA) of National Taiwan University (103R4000).
		
		\clearpage
		
		\appendix\label{app}
		\section{Details on the consistency check}
		\begin{subequations}
			
			Let us choose the natural units \(G=\hbar=c=1\). Here we try to argue that both Euclidean \cite{GMW} and Hamiltonian \cite{FMP} approaches to the tunneling process lead to the same physical result, decay rate. In our case, we analyze a \textit{true} bubble decay process with external cosmology constant zero, \textit{negative} internal cosmology constant \(
			\LG_I\), and the shell energy density per unit area is \(\m/4\pi\) which also stands for the surface tension \(\sigma={\m}/{4\pi}\). For more general cases, we can always consider zero or negative \(\LG_I\) as well as positive. 
			
			On the other hand, to apply thin-shell approximation, we assume the bubble shell to be spherical with negligible thickness. Since the Birkhoff's theorem guarantees that a spherically symmetry gravitational field has no dynamical degree of freedom, one can simplify the situation into a system with single degree of freedom, the bubble radius \(r\). In this way, our potential can be also simplified into the effective potential \(V(r)\) in \eqref{EOM}.
			
			Let's begin with the integration from the Euclidean action gives the following form \cite{PYD}
			\bea\label{1a}
			\int drr\left[\cos^{-1}\left(\frac{\bG_+}{\sqrt{f_+}}\right)-\cos^{-1}\left(\frac{\bG_-}{\sqrt{f_-}}\right)\right],
			\eea
			where extrinsic curvatures can be expressed as
			\bea\label{1b}
			\bG_\pm=\frac{f_--f_+\mp16\pi^2\sigma^2r^2}{8\pi\sigma r}.
			\eea
			By using the identity \[\cos^{-1}\aG-\cos^{-1}\bG=\cos^{-1}\big[\aG\bG+\sqrt{(1-\aG^2)(1-\bG^2)}\big],\] this is consistent with the second term in \eqref{tunnelingexponent}.
			
			In the Hamiltonian tunneling exponent\cite{FMP}, the \textit{shell-integration} is given as 
			\bea\label{1c}
			\hat{F}[r_2-r_1] 
			\propto \int_{r_1}^{r_2}drr
			\left[\cos^{-1}\left(\frac{2M-\m^2r^3(\lG-1)}{2\m r^2\sqrt{1-\frac{\LG_Ir^2}{3}}}\right)-
			\cos^{-1}\left(\frac{2M-\m^2r^3(\lG+1)}{2\m R^2\sqrt{1-\frac{2M}{r}}}\right)\right],
			\eea
			where \(\lG \equiv \LG_I/3\m^2\). 
			What we are trying to do is to make sure \eqref{1a} and \eqref{1c} are equivalent. In FMP \cite{FMP}, the authors solved the constraints for the momenta and identified the matching conditions by considering the geometry at the shell is smooth. Then they simplified results of constraints into a two-component vector which has a norm as,    
			\bea
			\textbf{V}\cdot\textbf{V}|_{internal}&=&1-\frac{\LG_I r^2}{3},\\
			\textbf{V}\cdot\textbf{V}|_{external}&=&1-\frac{2M}{r},
			\eea 
			which obeys the $SO(1,1)$ symmetry which is independent of angles. With a careful look, one can identify the notations of FMP as follows:
			\bea
			f_+&=&1-\frac{2M}{r}=\textbf{V}\cdot\textbf{V}|_{external}=1-\frac{2M}{r},\quad M\equiv M,\\
			f_-&=&1+\frac{r^2}{\ell^2}=\textbf{V}\cdot\textbf{V}|_{internal}=1-\frac{\LG_Ir^2}{3},\quad \ell^2\equiv-\frac{\LG_I}{3},
			\eea
			where the Schwarzschild radius is \(r_s=2M\) and the de Sitter radius is \(r_d=\sqrt{{3}/{\LG_I}}\). Thus from \eqref{1a}, together with \(\sigma={\m}/{4\pi}\) and \(\lG={\LG_I}/{3\m^2}\),
			\bea
			\frac{\bG_+}{\sqrt{f_+}}&=&\frac{\frac{f_--f_+-16\pi^2\sigma^2r^2}{8\pi \sigma r}}{\sqrt{1-\frac{2M}{r}}}
			=\frac{(1-\frac{\LG_Ir^2}{3})-(1-\frac{2M}{r})-16\pi^2(\frac{\m}{4\pi})^2r^2}{8\pi \sigma r(\frac{\m}{4\pi}) \sqrt{1-\frac{2M}{r}}}\\
			&=&\frac{2M-\m^2r^3\LG-\m^2r^3}{2\m  r^2\sqrt{1-\frac{2M}{r}}}=\frac{2M-\m^2r^3(\lG+1)}{2\m  r^2\sqrt{1-\frac{2M}{r}}}.
			\eea
			Similarly,
			\bea
			\frac{\bG_-}{\sqrt{f_-}}=\frac{2M-\m^2r^3(\lG-1)}{2\m r^2\sqrt{1-\frac{\LG_Ir^2}{3}}}.
			\eea
			Thus, we have safely connected \eqref{1a} and \eqref{1c}.
			
		\end{subequations}

		\bibliographystyle{plain}
		
\end{document}